\documentclass[usenatbib]{mn2e}
\usepackage[letterpaper,margin=0.8in]{geometry}
\usepackage{latexsym,graphicx,rotating,amsmath,amssymb, epsfig,hyperref}
\usepackage{astrojournals}
\usepackage[usenames, dvipsnames]{xcolor}

\newcommand\Beq{\begin{eqnarray}} 
\newcommand\Eeq{\end{eqnarray}}

\renewcommand{\vec}[1]{\boldsymbol{#1}}

\renewcommand{\dot}{\vec{\cdot}}

\newcommand{\grad}{\vec{\nabla}}

\renewcommand{\perp}{\!\bot}

\begin{document}

\title[Surface Manifestation of Stochastic IGWs]{Surface Manifestation of Stochastically Excited Internal Gravity Waves}
\author[Lecoanet et al]{Daniel Lecoanet$^{1,2}$\thanks{daniel.lecoanet@northwestern.edu}, Matteo Cantiello$^{3,4}$, Evan H.~Anders$^{2}$, Eliot Quataert$^4$,
\newauthor Louis-Alexandre Couston$^5$, Mathieu Bouffard$^6$, Benjamin Favier$^6$, \& Michael Le Bars$^6$ \\
$^1$ Department of Engineering Sciences and Applied Mathematics, Northwestern University, Evanston IL 60208, USA \\
$^2$CIERA, Northwestern University, Evanston IL 60201, USA \\
$^3$Center for Computational Astrophysics, Flatiron Institute, New York, NY 10010, USA \\
$^4$Department of Astrophysical Sciences, Princeton Univesity, Princeton, NJ 08544, USA \\
$^5$Univ Lyon, ENS de Lyon, Univ Claude Bernard, CNRS, Laboratoire de Physique, F-69342 Lyon, France \\
$^6$Aix Marseille Universit\'e, CNRS, Centrale Marseille, IRPHE, Marseille, France}

\maketitle

\begin{abstract}
Recent photometric observations of massive stars show ubiquitous low-frequency ``red-noise'' variability, which has been interpreted as internal gravity waves (IGWs).
Simulations of IGWs generated by convection show smooth surface wave spectra, qualitatively matching the observed red-noise.
On the other hand, theoretical calculations by Shiode et al (2013) and Lecoanet et al (2019) predict IGWs should manifest at the surface as regularly-spaced peaks associated with standing g-modes.
In this work, we compare these theoretical approaches to simplified 2D numerical simulations.
The simulations show g-mode peaks at their surface, and are in good agreement with Lecoanet et al (2019).
The amplitude estimates of Shiode et al (2013) did not take into account the finite width of the g-mode peaks; after correcting for this finite width, we find good agreement with simulations.
However, simulations need to be run for hundreds of convection turnover times for the peaks to become visible; this is a long time to run a simulation, but a short time in the life of a star.
The final spectrum can be predicted by calculating the wave energy flux spectrum in much shorter simulations, and then either applying the theory of Shiode et al (2013) or Lecoanet et al (2019).
\end{abstract}

\section{Introduction}

The detection of ubiquitous low-frequency variability in massive stars by \citet{Bowman2019} has revealed a potential keyhole through which we may better understand massive stellar structure and evolution.
Massive stars play an important role in astrophysics, and are progenitors for compact binary systems whose mergers produce gravitational waves.
The successful interpretation of the low-frequency variability in massive stars could answer important questions about the age, angular momentum transport, mass-loss history, and chemical mixing in massive stars \citep{Bowman2019}.

There are two main physical interpretations of the low-frequency variability: internal gravity waves \citep[e.g.,][]{Bowman2019,Ratnasingam2020,Bowman2020,Horst2020}; and (sub)surface convection \citep[e.g.,][]{Cantiello2021}.
Massive stars have convective cores which can generate internal gravity waves, which subsequently travel to the surface of the star.
In previous work, we suggested that the surface frequency spectrum of these waves would be dominated by regularly-spaced standing mode peaks \citep{Lecoanet2019}, inconsistent with the observations of relatively smooth profiles without clearly identifiable features such as peaks \citet{Bowman2019}.
The wave-forcing simulations of \citet{Ratnasingam2020} also show surface frequency spectra dominated by peaks, especially for simulations of stars that evolved off the ZAMS.
These results seem to contradict many simulations which show smooth wave frequency spectra near the surface of massive stars \citep[e.g.,][]{Rogers2013,Edelmann2019,Horst2020}.
Separately, \citet{Shiode2013} calculated the amplitude of standing g-modes in massive stars, and found the typical amplitude to be $\,\sim 10-100 \mu{\rm mag}$, much smaller than the typical observed low-frequency variability.

Up to now, studies of internal wave generation by convection have taken either a primarily quasi-analytical approach \citep[e.g.,][]{Shiode2013, Lecoanet2019}, or a primarily numerical approach \citep[e.g.,][]{Rogers2013, Edelmann2019, Horst2020}.
The goal of this paper is to bridge the gap.
We will demonstrate the theoretical predictions in \citet{Shiode2013} and \citet{Lecoanet2019} are in agreement with fully nonlinear numerical simulations.
One key aspect is that numerical simulations need to be run for hundreds of convection times to reach saturated wave amplitudes.
For this reason, we study a simplified 2D Boussinesq setup.
Future work will explore these issues in more realistic 3D spherical simulations.

\section{Simulation Setup}\label{sec:setup}

To determine the surface manifestation of convectively excited waves, we run a series of simple 2D Cartesian simulations.
We run 2D simulations similar to \citet{Couston2017,Couston2018b}; we restrict ourselves to 2D because this work requires very long integrations.
In this model, we solve the Boussinesq equations with a piece-wise linear equation of state,
\begin{align}\label{eqn:momentum}
\partial_t\vec{u}+\grad p - \nu\nabla^2 \vec{u} &= -\vec{u}\dot\grad\vec{u} - g\frac{\delta\rho}{\rho_0}\vec{e}_z, \\
\grad\dot\vec{u} &= 0, \\
\partial_t T -\kappa\nabla^2 T &= -\vec{u}\dot\grad T, \label{eqn:temperature}
\end{align}
where $\vec{u}$ and $p$ are the velocity, pressure, $T$ is the temperature perturbation, $\nu$ and $\kappa$ are the viscosity and thermal diffusivity, $g$ is the strength of gravity, $\vec{e}_z$ is the unit vector in the vertical direction.
The density perturbation is given by
\begin{align}
\frac{\delta\rho}{\rho_0}=
\begin{cases}
-\alpha T & {\rm for}\ T > 0 \\
\alpha S T & {\rm for}\ T < 0
\end{cases},
\end{align}
where $\alpha$ is the coefficient of thermal expansion, and $S$ is the stiffness parameter.
The temperature perturbation $T$ is defined to be zero at the temperature of the density maximum.

We solve these equations on a domain with length $L=1$ in both the horizontal $x$ direction, and the vertical $z$ direction.
We refer to the top of the domain ($z=1$) as the ``surface'' of the simulation, in analogy to a simulation of a star.
The horizontal boundary conditions are periodic, and the vertical boundary conditions are stress-free and fixed temperature.
The temperature perturbation is fixed to $T=1$ at the bottom boundary and $T=T_{\rm top} < 0$ at the top boundary.
In the bottom part of the domain, $T>0$, so the density perturbation is given by $-\alpha T$, and the fluid is unstable to convection; whereas in the top part of the domain, $T<0$, so the density perturbation is given by $\alpha S T$, and the fluid is stably stratified.
The convection is driven by an unstable temperature jump of $\Delta T= 1$.
In this model, the radiative-convective boundary (corresponding to $T=0$) is determined self-consistently.
We pick $T_{\rm top}$ such that the height of the convection and the height of the radiative zone are both about $L/2$.
The Brunt-V\"{a}is\"{a}l\"{a} frequency is given by $N^2 = g \alpha S (dT/dz)$.
Thus, larger values of the stiffness correspond to larger values of $N$.
In this work, we use $S=100$, which is the high-stiffness regime where the convection is only weakly modified by the presence of the radiative zone \citep{Couston2017}.
The remaining parameters are chosen such that the convective buoyancy timescale $\sqrt{g\alpha\Delta T/L}=1$.

\begin{table*}
\centering
\begin{tabular}{cccrcccrr}\hline
Name & $Ra$ & damping & $T_{\rm top}$ & $N_x\times(N_{z,c}+N_{z,r})$ & $\tau_c$ & $\tau_c N/(2\pi)$ & $t_{\rm sim}/\tau_c$ & $t_a/\tau_c$ \\ \hline \hline
$C^{8}$ & $2\times 10^{8}$ & no & $-60$ & $512 \times (512 + 256)$ & 1.19 & 20.7 & 1179 & 1166 \\
$D^{8}$ & $2\times 10^{8}$ & yes & $-60$ & $512 \times (512 + 256)$ & 1.35 & 23.5 & 70 & 59 \\
$C^{9}$ & $10^{9}$ & no & $-100$ & $1024\times (1024 + 512)$ & 1.27 & 28.6 & 333 & 85 \\
$D^{9}$ & $10^{9}$ & yes & $-100$ & $1024 \times (1024 + 512)$ & 1.53 & 34.4 & 44 & 34 \\
$C^{10}$ & $10^{10}$ & no & $-200$ & $1536 \times (1536 + 768)$ & 1.30 & 41.4 & 349 & 165 \\
$D^{10}$ & $10^{10}$ & yes & $-200$ & $1536 \times (1536 + 768)$ & 1.73 & 55.0 & 25 & 18 \\
\end{tabular}
\caption{Simulations described in this paper. $Ra$ is the Rayleigh number.
$T_{\rm top}$ is the temperature perturbation of the top boundary, set such that the radiative-convective boundary is close to $z=0.5$.
The number of horizontal Fourier modes is $N_x$, and the number of vertical Chebyshev modes in the convection and radiative zones are $N_{z,c}$ and $N_{z,r}$, respectively.
The convection timescale is defined as $\tau_c=0.5/u_{\rm rms}$, and we measure times in units where the convective buoyancy time $\sqrt{g \alpha\Delta T/L}=1$.
We also report the ratio of the convective timescale to the buoyancy timescale, $2\pi/N$.
The simulations were run for a total time $t_{\rm sim}$, and simulation analysis occurs over the time $t_a$.}\label{tab:sims}
\end{table*}

The level of turbulence in the convection zone can be parameterized by the Rayleigh number, which is the ratio of convective driving (given by the unstable temperature gradient when $T>0$), to diffusive damping (given by the product of the diffusivities),
\begin{equation}
Ra = \frac{g\alpha \Delta T (L/2)^3}{\nu\kappa} = \frac{1}{8\nu\kappa}.
\end{equation}
Note we use the height of the convection zone, $L/2$, as the relevant lengthscale.
All our simulations have Prandtl number unity, so $\nu=\kappa$.
As the Rayleigh number increases, the convection becomes more turbulent, and the waves also experience less damping in the radiative zone.
This leads to lower-frequency waves at the surface, as well as narrower standing-mode peaks.

We solve the equations using the Dedalus pseudo-spectral code \citep{Burns2020}.
Variables are represented as Fourier series in the horizontal $x$ direction with $N_x$ Fourier modes.
In the vertical $z$ direction, we represent each variable using one set of Chebyshev polynomials for the interval $0\leq z \leq z_{\rm int}=0.6$, and another set of Chebyshev polynomials for the interval $z_{\rm int} \leq z \leq 1$.
We use $N_{z,c}$ Chebyshev polynomials for the lower interval in the convection zone, as we use $N_{z,r}$ Chebyshev polynomials for the upper interval in the radiative zone.
At $z=z_{\rm int}$, we impose continuity of all variables.
We use this ``matched-Chebyshev'' discretization to increase the vertical resolution of our simulation near the radiative--convective boundary at $z=1/2$.
To avoid aliasing errors, we use the 3/2-dealiasing rule in both horizontal and vertical directions.
For timestepping, we use a 2nd-order, two-stage, implicit-explicit Runge-Kutta scheme \citep{Ascher97}, where all linear terms are treated implicitly, and all nonlinear terms are treated explicitly (including the buoyancy term).
The timestep size is chosen according to the CFL criterion, which is applied only for $z<0.52$, and with a safety factor of 0.35.
We only apply the CFL criterion below $0.52$ because the grid-spacing becomes extremely fine near $z_{\rm int}$, and we have found that the accurate propagation of internal gravity waves does not require us to satisfy the CFL criterion.

We run two types of simulations.
In the first type, we include an additional damping term, $-\vec{u} h(z)/\tau$, to the right hand side of the velocity equation.
The vertical structure of the damping layer is given by $h(z)=(1+\tanh\left[(z-0.925)/(0.025)\right])/2$, so that it damps out waves in the region $z\gtrsim 0.925$.
The damping time is given by $\tau^{-1}=30$ (recall we measure times in units where the convective buoyancy timescale $\sqrt{g\alpha\Delta T/L}=1$).
In these simulations, vertically propagating waves are mostly damped out by this layer, and there is little reflection.
The second type of simulation does not include the damping layer, so waves reflect off the top boundary.
This leads to resonances near the eigenfrequencies of the radiative zone.
Simulations are initialized with zero velocity, and
\begin{equation}
T = \left[1-z+\theta(z) - \theta(0)\right] + T_{\rm top}\left[z+\theta(z)-\theta(0)\right] + \mathcal{N},
\end{equation}
where $\mathcal{N}$ is low-amplitude random noise in the convection zone ($z<0.4$), and
\begin{equation}
\theta(z) = \Delta z \log\left[\cosh\left(\frac{z-0.5}{\Delta z}\right)\right],
\end{equation}
with $\Delta z = 0.02$.
The temperature perturbation smoothly transitions from one linear curve from $T(0)=1$ to $T(0.5)=0$, and a second linear curve from $T(0.5)=0$ to $T(1)=T_{\rm top}<0$.
All simulation, analysis, and plotting scripts can be found at \url{https://github.com/lecoanet/2D_waveconv}.

The parameters of all our simulations are described in Table~\ref{tab:sims}.
We also report the convection time $\tau_c$ and the total simulation length $t_{\rm sim}$.
The convection time is estimated using $\tau_c = 0.5/u_{\rm rms}$, where $u_{\rm rms}^2$ is the temporal and horizontal average of $|\vec{u}|^2$ at $z=0.4$, which captures the typical velocities in the bulk of the convection zone.
The simulations are run for a total time $t_{\rm sim}$, but all the analysis presented in this work (including calculating $u_{\rm rms}$) is performed over a shorter analysis time window, $t_a$, which avoid transients.

\section{Wave Flux Spectra in Simulations with Damping}\label{sec:damping}

\begin{figure}
  \centering
  \includegraphics[width=\linewidth]{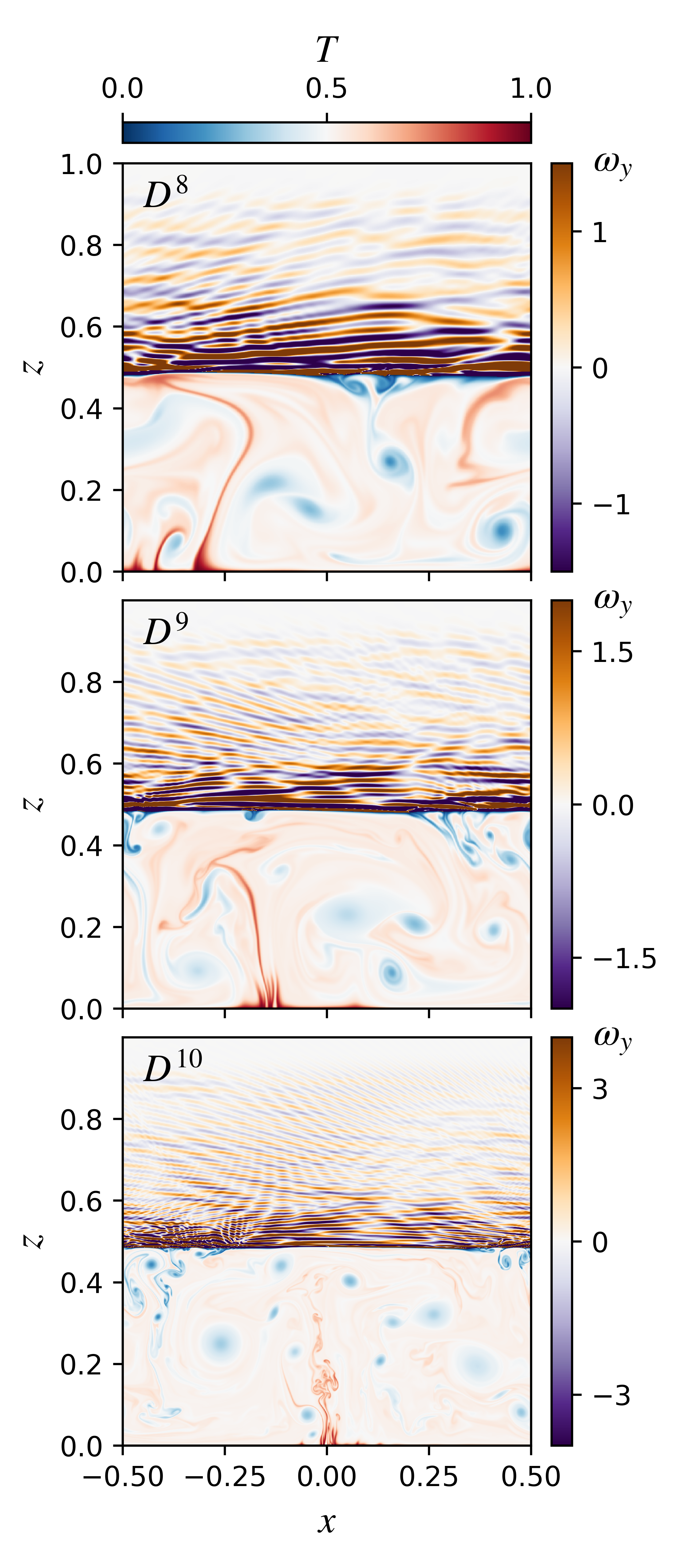}
  \caption{Visualization of simulations with damping layers: $D^8$, $D^9$, and $D^{10}$. The simulations without damping layers look similar, but the waves extend to the top of the domain.
  For $T>0$, we plot the temperature perturbation to visualize the convection, and for $T<0$, we plot the vorticity, $\omega_y$, to visualize the waves.
  The simulations with higher Rayleigh number and lower diffusivities show finer-scale structures in both the convection zone and the radiative zone.
  As the waves propagate upward, the lower-frequency, mostly horizontal waves are preferentially damped, leaving mostly high-frequency waves near the top of the domain.}
  \label{fig:convection}
\end{figure}

\begin{figure*}
  \centering
  \includegraphics[width=\linewidth]{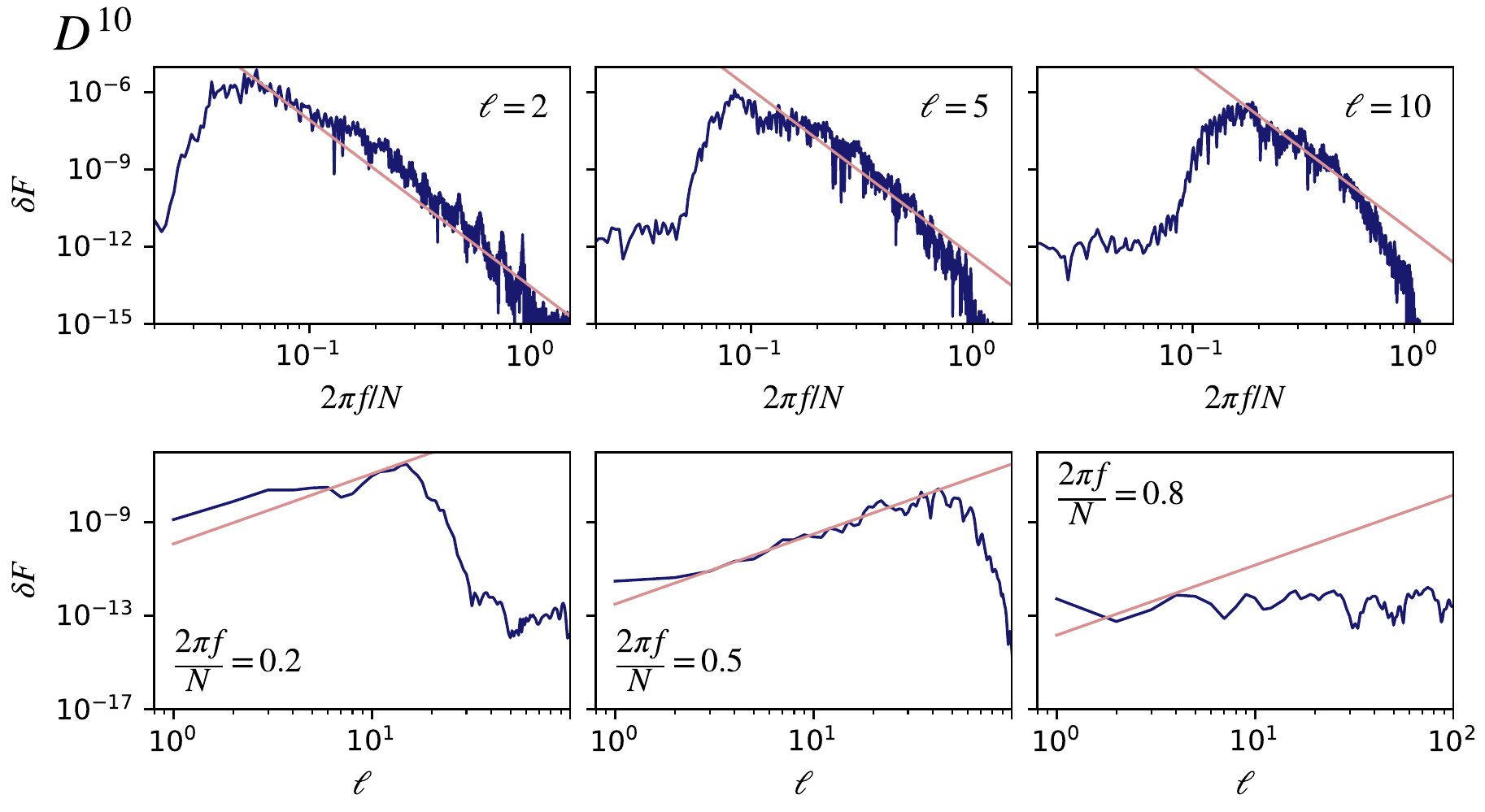}
  \caption{The wave flux spectrum as a function of $f$ (top row) and $\ell=k_x/(2\pi)$ (bottom row) for simulation $D^{10}$. The dark blue curve shows the simulation data; the light red curve shows the power-law expression using $\mathcal{A}=2\times10^{-5}$, $a=3$, and $b=-13/2$ (see equation~\ref{eqn:fit}). We plot the wave flux and the power-law for three illustrative values of $\ell$ and $f$. The power-law expression works well, except for the highest frequency we plot here (lower right plot). Appendix~\ref{sec:waveflux89} includes similar plots for simulations $D^8$ and $D^9$.}
  \label{fig:waveflux}
\end{figure*}

We start by describing the simulations $D^8$, $D^9$, and $D^{10}$, which all include a damping layer at the top of the simulation domain.
This damping layer inhibits wave reflection, which makes them useful for measuring the wave energy flux.
The wave energy flux can be used to predict the surface manifestation of internal gravity waves, as described in section~\ref{sec:surface}.
The wave energy flux equilibrates much more quickly than the frequency spectrum of perturbations at the top boundary, which we measure in our simulations without a damping layer.
For this reason, the simulations with a damping layer are run for a much shorter time than the simulations without a damping layer.
There is a slow thermal equilibration in our simulations as the size of the convection zone slowly adjusts.
The convective velocities slowly change over time, causing $\tau_c$ to be somewhat different in the simulations with and without a damping layer.
Because we measure time in units where the convective buoyancy timescale is 1, $\tau_c$ is never very different from 1.

Figure~\ref{fig:convection} shows snapshots from all three simulations with damping layers near the end of the simulation.
In the convection zone (where $T>0$), we plot the temperature perturbations, which allows one to visualize warm and cold plumes.
Because our simulations are two-dimensional, they are dominated by vortices.
As the Rayleigh number increases, the size of the vortices decreases, and the convective plumes break up into a string of vortices \citep{Johnston2009, Zhu2018}.
In the radiative zone (where $T<0$), the temperature fluctuations are small, and the temperature decreases roughly linearly from $T=0$ at $z\approx 0.5$ to $T=T_{\rm top}<0$ at $z=1$.
To visualize waves, we plot the vorticity $\omega_y=\partial_z u_x - \partial_x u_z$, where $u_x$ and $u_z$ are the horizontal and vertical velocities.
The magnitude of the vorticity decreases with height because the waves experience damping as they propagate upward.
There is very little vorticity in the damping region $z>0.925$.
Near the radiative-convective interface, the waves are predominantly horizontal: the convection is most efficient at exciting waves near the convective frequency, which is much smaller than $N$ (see table~\ref{tab:sims}).
However, these low-frequency waves damp out very quickly.
The only waves that can successfully propagate toward the top of the domain are higher-frequency waves with frequencies closer to $N$, which are less horizontal.

We quantitatively characterize the simulations with a damping layer by calculating the wave energy flux at the height $z=0.6$.
The wave energy flux is $F=u_zp$, where $u_z$ is the vertical velocity and $p$ is the pressure. 
$F$ is a useful quantity because, neglecting diffusive effects, it is conserved for linear waves.
\citet{Lecoanet2013} extended the results of \citet{Goldreich1990} to make a theoretical prediction for the wave flux,
\begin{align}\label{eqn:LQ13}
\frac{dF}{d k_{\perp} df} = \mathcal{A}_c \, \frac{L \tau_c}{2\pi} \left(\frac{k_{\perp} L}{2\pi}\right)^3 (f\tau_c)^{-15/2},
\end{align}
where $f$ is the wave frequency, $k_{\perp}$ is the wavenumber perpendicular to gravity, and the coefficient $\mathcal{A}_c$ is predicted to scale like $F_c/(\tau_cN)$, where $F_c$ is the convective flux.
The power-law form is only valid for $k_{\perp}L\lesssim (f\tau_c)^{3/2}$ and $f \gtrsim \tau_c^{-1}$, so it does not diverge at large wavenumbers or low frequencies.
This theoretical prediction is in good agreement with wave flux spectra measured in Boussinesq simulations of wave generation by convection in 3D Cartesian domains \citep{Couston2018}.
The prediction is derived by assuming that turbulent convection can be decomposed into eddies of different sizes, and each eddy has an amplitude given by the $E(k)\sim k^{-5/3}$ Kolmogorov spectrum, and is coherent for its turnover time.
In our simulations, we find a steep $E(k) \sim k^{-4}$ spectrum for $k/(2\pi) \gtrsim 10$, consistent with experiments and simulations of forced 2D turbulence \citep{Boffetta2012}.
Thus, one would expect wave flux spectra measured in 2D numerical simulations to not agree with equation~\ref{eqn:LQ13}.

We calculate the wave energy flux by taking the spatial and temporal Fourier transforms of $u_z$ and $p$.
We normalize the Fourier transforms such that
\begin{align}
\frac{1}{t_a} \int_{t_0}^{t_0+t_a} u_z^2 dt = \sum_f |\hat{u}_z(f)|^2,
\end{align}
where $\hat{u}_z$ is temporal Fourier transform of $u_z$, and we analyze the data from $t_0$ to $t_0+t_a$.
When calculating frequency spectra, we first multiply the timeseries by a Hann function.
We use similar relations for the horizontal Fourier transform and for defining $\hat{p}$.
Hereafter, we will use $\hat{\cdot}$ to mean the horizontal and temporal Fourier transform of a variable.
Then the wave flux is given by
\begin{align}
\frac{1}{t_a} \int_0^1 dx \int_{t_0}^{t_0+t_a} dt \, u_z p &= \sum_{k_x, f} \Re\left[\hat{u}_z \hat{p}^*\right] \nonumber \\
&= \sum_{k_x,f} \delta F,
\end{align}
and the differential wave flux is given by
\begin{align}
\frac{dF}{dk_x df} = \frac{\delta F}{\delta k_x \delta f} = \frac{t_a}{2\pi}\Re\left[\hat{u}_z \hat{p}^*\right].
\end{align}
To simplify notation, we define
\begin{align}
\ell = \frac{k_x}{2\pi},
\end{align}
such that $\ell=1$ corresponds to a wave at the domain size.

We parameterize the wave flux spectra in our simulations with a damping layer using the power-law form
\begin{align}\label{eqn:fit}
\delta F = \mathcal{A} \, \ell^a f^{b},
\end{align}
where we allow $\mathcal{A}$ and the power-law exponents $a$ and $b$ to vary.
We plot the wave flux spectrum of simulation $D^{10}$ in figure~\ref{fig:waveflux}, and also include the flux spectra of simulations $D^8$ and $D^9$ in appendix~\ref{sec:waveflux89}.

When $\ell=k_x/(2\pi)$ is fixed, we find that the wave flux decreases like $f^{-13/2}$.
This is similar to, but does not exactly match, the prediction in equation~\ref{eqn:LQ13}.
Figure~\ref{fig:waveflux} also shows some weak peaks in the spectrum at frequencies associated with the eigenfrequencies of the radiative zone; however, these are much lower amplitude than they would be if there was no damping layer (see figure~\ref{fig:spectrum_ell1}).
Power-laws with $b=-13/2$ or $b=-15/2$ seem to give a good match to all three simulations.
For the wavenumber dependence, we find the wave flux increases with increasing wavenumber until a critical value, and then decreases abruptly.
All three simulations are consistent with a wavenumber power-law with $a=3$ (except maybe at high frequencies), in agreement with equation~\ref{eqn:LQ13}.
In all the simulations, it is difficult to exactly determine the power-law indices $a$ and $b$.
We picked indices that seemed consistent with the data, were similar to the prediction of equation~\ref{eqn:LQ13}, and which matched the surface spectra as measured in figures~\ref{fig:spectrum_all} \& \ref{fig:spectrum_freq}.
The amplitudes $\mathcal{A}$ were then chosen to match the wave flux data.
Table~\ref{tab:waveflux} lists the parameters we use to match the simulations with damping layers.
Overall, we find there is unexpectedly good agreement with \citet{Lecoanet2013}.
This is surprising, as the theory should not be applicable for these 2D simulations.
This suggests there may be an alternative explanation for equation~\ref{eqn:LQ13} that would lead to the same prediction, but also would be applicable to these 2D simulations.

\begin{table}
\centering
\begin{tabular}{cccc}\hline
Name & $\mathcal{A}$ & $a$ & $b$ \\ \hline \hline
$D^{8}$ & $10^{-5}$ & 3 & $-15/2$ \\
$D^{9}$ & $2\times10^{-6}$ & 3 & $-13/2$ \\
$D^{10}$ & $2\times10^{-5}$ & 3 & $-13/2$ \\
\end{tabular}
\caption{Wave flux spectrum parameters for the simulations with damping layers. The parameters are specified in equation~\ref{eqn:fit}.}\label{tab:waveflux}
\end{table}

\section{Surface Manifestation of Internal Gravity Waves}\label{sec:surface}

\begin{figure}
  \centering
  \includegraphics[width=\linewidth]{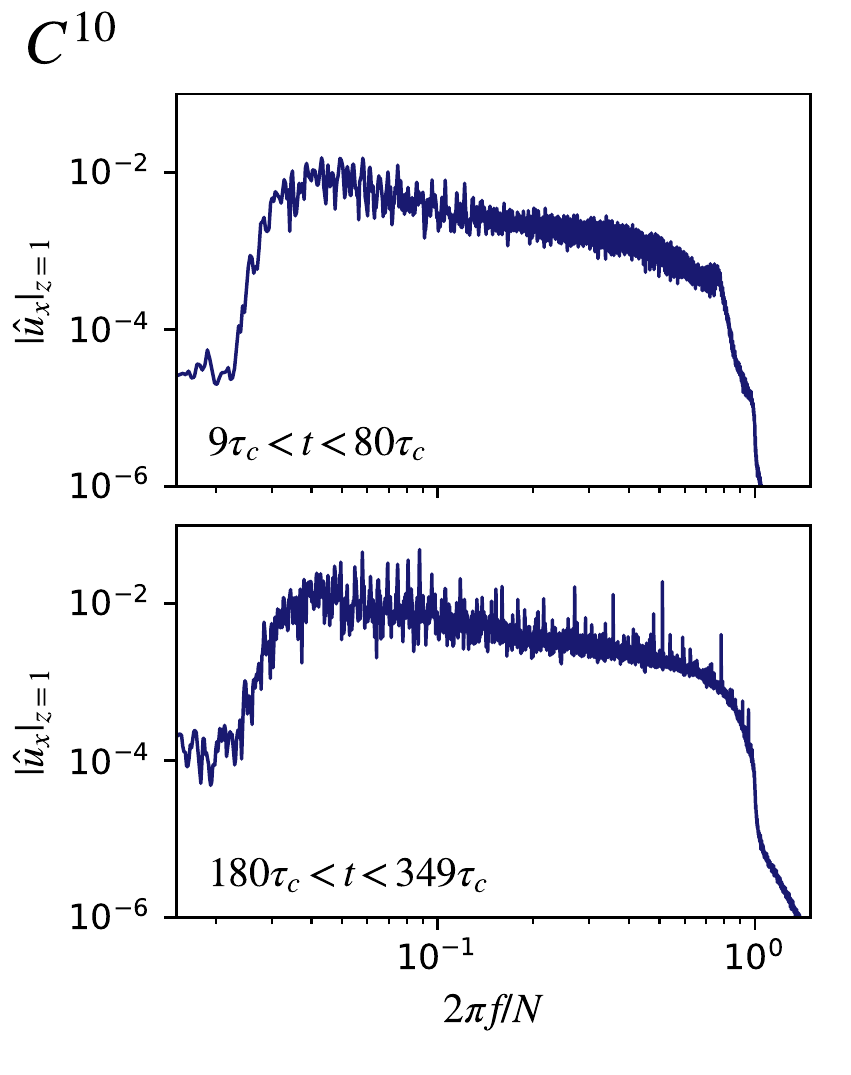}
  \caption{The frequency spectrum of the horizontal velocity $u_x$ at the ``surface'' of simulation $C^{10}$, at $z=1$. The top panel shows the spectrum early in the simulation, and the bottom panel shows the spectrum late in the simulation (extending to $\approx 350$ convection times). It takes more than 100 convection times before sharp peaks associated with the radiative zone's eigenfrequencies become apparent.}
  \label{fig:spectrum_C10}
\end{figure}

\begin{figure}
  \centering
  \includegraphics[width=\linewidth]{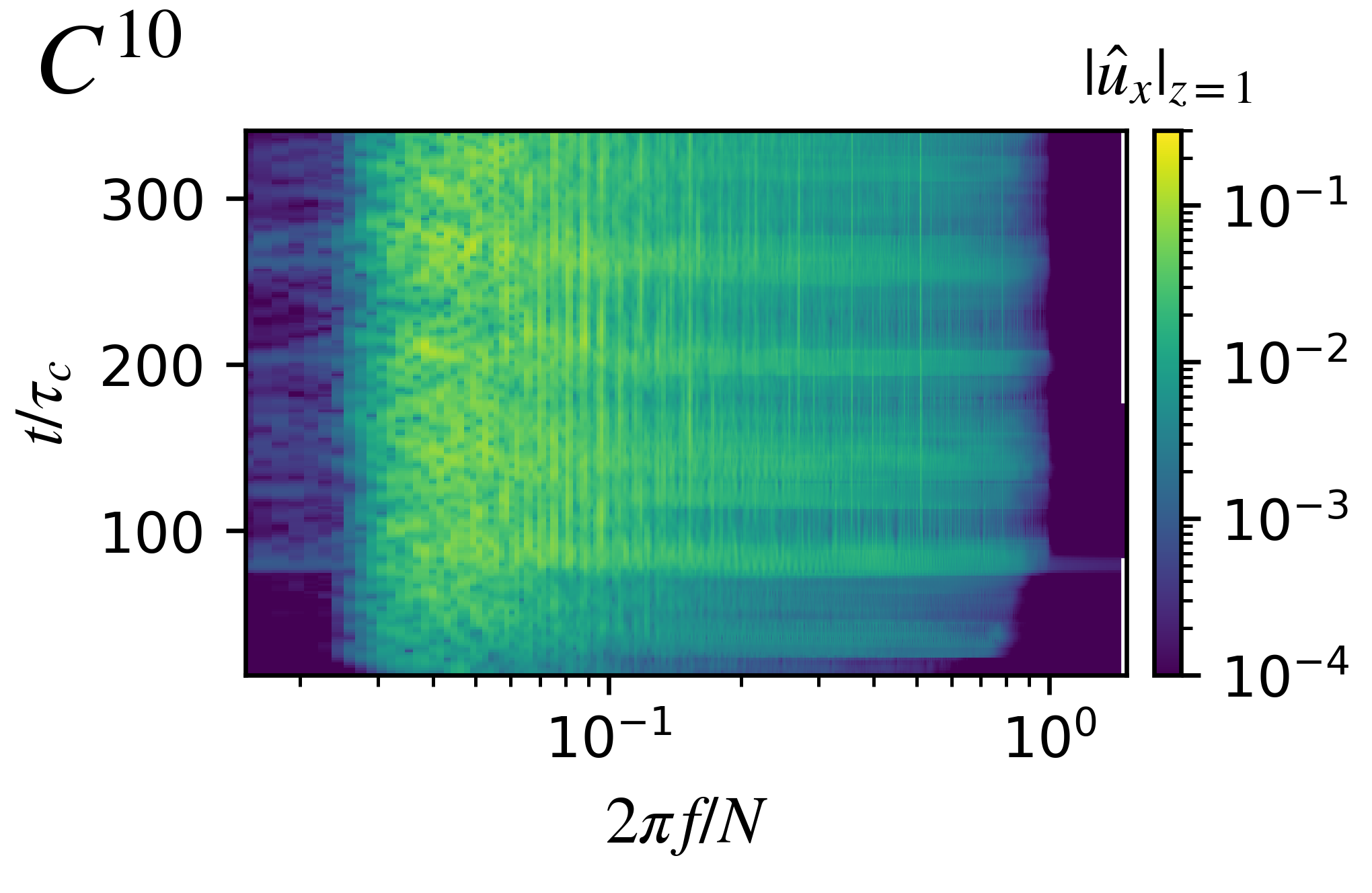}
  \caption{Frequency spectrum of the horizontal velocity at $z=1$ in simulation $C^{10}$ as a function of time.
  The white bars on the right side of the plot show the averaging windows of the spectra in figure~\ref{fig:spectrum_C10}.
  It takes a long time ($>100\tau_c$) for standing modes (bright vertical lines) to develop.
  Intense convective events generate stronger waves, and appear as horizontal stripes on the plot.}
  \label{fig:spectrogram}
\end{figure}

We now analyze the simulations without a damping layer.
In these simulations, the vertical velocity and temperature perturbations at the top boundary are both fixed to constants.
We use stress-free boundary conditions ($\partial_zu_x=0$), so we measure the surface manifestation of waves by measuring the horizontal velocity spectrum at the top boundary.
Although we focus on horizontal velocity here, we could have used any non-zero wave perturbation variable.
Thus, we believe this analysis will also carry over to calculating the surface luminosity perturbation in a star.

The main goal of this paper is to predict the frequency spectrum of $u_x$ at the ``surface'' of the simulation, i.e., at $z=1$.
One difficulty is that the surface spectrum evolves secularly over time.
In figure~\ref{fig:spectrum_C10}, we plot the surface spectrum for our high-resolution simulation $C^{10}$ at early times ($9\tau_c < t < 80\tau_c$), and at late times ($180\tau_c< t < 349\tau_c$).
At early times, the spectrum is relatively smooth with a broad maximum near $2\pi f\sim 0.04 N$, but at late times, the spectrum shows many sharp peaks associated with the eigenfrequencies of the radiative zone.
These peaks are clear features in the surface spectra of all our simulations after sufficient temporal integration.

In Figure~\ref{fig:spectrogram} we plot frequency spectra calculated over short time windows, to visualize how the wave amplitudes change with time.
Each horizontal line in the figure shows the spectrum calculated between $t-\Delta t/2$ and $t+\Delta t/2$, where the time window is $\Delta t\approx 15 \tau_c$.
We vary the central time $t$ in increments of $\approx 1.5\tau_c$ to form the figure.
The wave amplitudes grow slowly at the beginning of the simulation.
Then an intense convection event at $t\approx 70\tau_c$ is able to produce strong waves, which appears as a bright horizontal band on the figure.
This starts to generate standing mode peaks (bright vertical bands), which continue to grow in amplitude with subsequent intense convection events.
Although the wave generation is strongly intermittent, the spectrum itself appears to be relatively steady in the last half of the simulation.
It is our goal to describe this statistically steady state.

The frequency spectra in figures~\ref{fig:spectrum_C10} \& \ref{fig:spectrogram} have contributions from many horizontal wavenumbers.
It is simpler to consider a single horizontal wavenumber at a time.
In figure~\ref{fig:spectrum_ell1} we plot the frequency spectrum of the $\ell=1$ component of $|\hat{u}_x|_{z=L}$ in simulation $C^{10}$.
There are many sharp peaks in the spectrum; these are at the eigenfrequencies of the radiative zone.
We also plot the predicted spectrum using a transfer function, as well as a second prediction for the mode amplitudes.
The two predictions match the simulated horizontal velocity spectrum.
We will now describe how these predictions are made.

\subsection{Wave Transfer Function}\label{sec:transfer}

First we will apply the transfer function approach of \citet{Lecoanet2019} to these Cartesian, Boussinesq simulations.
The main idea is to link the horizontal velocity at the surface to the vertical velocity near the radiative-convective boundary,
\begin{align}\label{eqn:transfer}
\hat{u}_x(z=1) = T(\ell, f) \hat{u}_z(z_{\rm RCB}).
\end{align}
For weakly damped waves, we calculate the transfer function $T(\ell, f)$ using an eigenfunction expansion.
First we calculate the eigenvalues and eigenfunctions according to appendix~\ref{sec:eigenvalue}.
We then calculate a dual basis to the eigenfunctions.
The dual basis, $\vec{u}_{n}^\dagger=(u_x^\dagger, u_z^\dagger)$ satisfies
\begin{align}
\left\langle \vec{u}_n^\dagger, \vec{u}_m\right\rangle = \delta_{n, m},
\end{align}
where $n$ and $m$ enumerate the eigenvalues, and the inner product is
\begin{align}
\left\langle \vec{f}, \vec{g}\right\rangle = \int_0^L \vec{g}^*\vec{\cdot}\vec{f} \, dz.
\end{align}
Following the appendix of \citet{Lecoanet2019}, we find that the transfer function is given by
\begin{align}\label{eqn:T}
T(\ell,f) = \frac{1}{\Delta z}\int_{z_{\rm RCB}}^{z_{\rm RCB}+\Delta z} dz_f \frac{i\sqrt{2} f}{\ell} Z(z_f; \ell, f).
\end{align}
We approximate the integral over $z_f$ by using 100 equally-spaced values of $z_f$ between $z_{\rm RCB}=0.5$ and $z_{\rm RCB}+\Delta z=0.6$; the transfer function is insensitive to the exact integration region used.

\begin{figure}
  \centering
  \includegraphics[width=\linewidth]{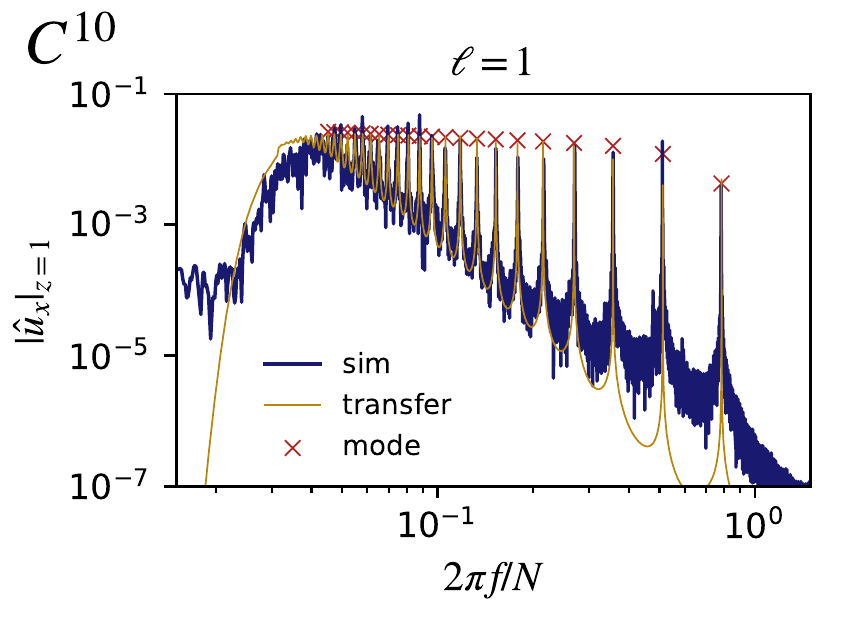}
  \caption{The frequency spectrum of the $\ell=1$ component of the horizontal velocity, $|\hat{u}_x|_{z=1}$. We also plot the predicted spectrum derived using a transfer function (thin yellow, equation~\ref{eqn:transfer fit}), as well as predicted mode amplitudes (red crosses, equation~\ref{eqn:mode}).}
  \label{fig:spectrum_ell1}
\end{figure}

The $Z$ function represents the eigenfunction expansion, where $\omega$ are the eigenfrequencies as derived in appendix~\ref{sec:eigenvalue},
\begin{align}\label{eqn:eigenfunction}
Z(z_f; \ell, f) = \sum_{\omega} \frac{u_x^\dagger(z_f;\ell, \omega) u_x(z=1;\ell, \omega)}{2\pi f-\omega}.
\end{align}
For the remainder of this paper, we use $f$ as the independent variable in our frequency spectra, and $\omega$ as an eigenvalue.
The eigenvalues $\omega$ are complex: the oscillation frequency is $\Re(\omega)/(2\pi)$ and the damping rate is $\gamma=\Im(\omega)$.
The transfer function at frequency $f$ is dominated by the eigenvalue $\omega$ closest to $2\pi f$.
There is an amplification of the surface manifestation of a mode by about $\Re(\omega)/\Im(\omega)$, which occurs when $2\pi f=\Re (\omega)$.
In simulation $C^{10}$, this ratio is $7\times 10^5$ for the highest-frequency, $\ell=1$ mode.
Figure~\ref{fig:spectrum_ell1} shows the transfer function amplifies by roughly this magnitude near $2\pi f\approx 0.8N$.
While the amplitude of the peak is similar in the simulation, the amplitude of the trough near this frequency is lower in the transfer function prediction than in the simulation.
The transfer function expression, equation~\ref{eqn:T}, is derived assuming the long-time limit, $t\gamma \gg 1$, where $\gamma$ is the damping rate of the mode $\omega$.
For this highest-frequency $\ell=1$ mode, it should take $1/\gamma$ time units to reach this amplitude, which corresponds to $\approx 3500 \tau_c$, a factor of 10 longer than our actual run time.
This is a possible explanation for the deviation of the simulations from the transfer function calculation at large $f/N$ in figure~\ref{fig:spectrum_ell1}.

The eigenfunction expansion of equation~\ref{eqn:eigenfunction} does not work well at low frequencies.
Eigenfunctions for this problem correspond to a superposition of upward-propagating and downward-propagating waves.
At low frequencies, the waves are strongly attenuated by diffusion, so very little power is reflected into downward-propagating waves.
This makes the eigenfunction expansion ill-suited for describing these low-frequencies waves.
To calculate the transfer function at low frequencies, we solved the linearized wave equations with a volumetric forcing term via direct time integration in Dedalus.
We run multiple simulations with different forcing frequencies, and with a forcing profile with width $\delta z=0.01$, centered at several locations $z_f$.
After the simulation reaches a statistically steady state, we measure $|u_x|_{z=1}$.
Then the function $Z$ is given by
\begin{align}\label{eqn:direct}
Z(\ell, f) = c\frac{|u_x|_{z=1}}{\delta z}
\end{align}
We expect the factor $c$ to be equal to unity.
However, we find that equations~\ref{eqn:eigenfunction} and \ref{eqn:direct} agree when we use $c=2$, so we use this value.
We describe the numerical details of this calculation in appendix~\ref{sec:forcing}.

Finally, in order to use equation~\ref{eqn:transfer} we need an expression for $\hat{u}_z(z_{\rm RCB})$.
We determine this using the wave energy flux.
We have
\begin{align}
\frac{1}{2}|\hat{u}_z|^2 = \frac{k_x}{N} \delta F,
\end{align}
where $\delta F$ is given by the power-law relation in equation~\ref{eqn:fit}.
We also adjust the overall amplitude slightly for each simulation with an amplitude factor $\mathcal{A}_T$, e.g., accounting for differences between the simulations with and without a damping layer.
The predicted wave flux using the transfer function is then
\begin{align}\label{eqn:transfer fit}
|\hat{u}_x|_{z=1} = \mathcal{A}_T \, \sqrt{\frac{2k_x\delta F}{N}} \, T(\ell, f).
\end{align}

\begin{figure*}
  \centering
  \includegraphics[width=\linewidth]{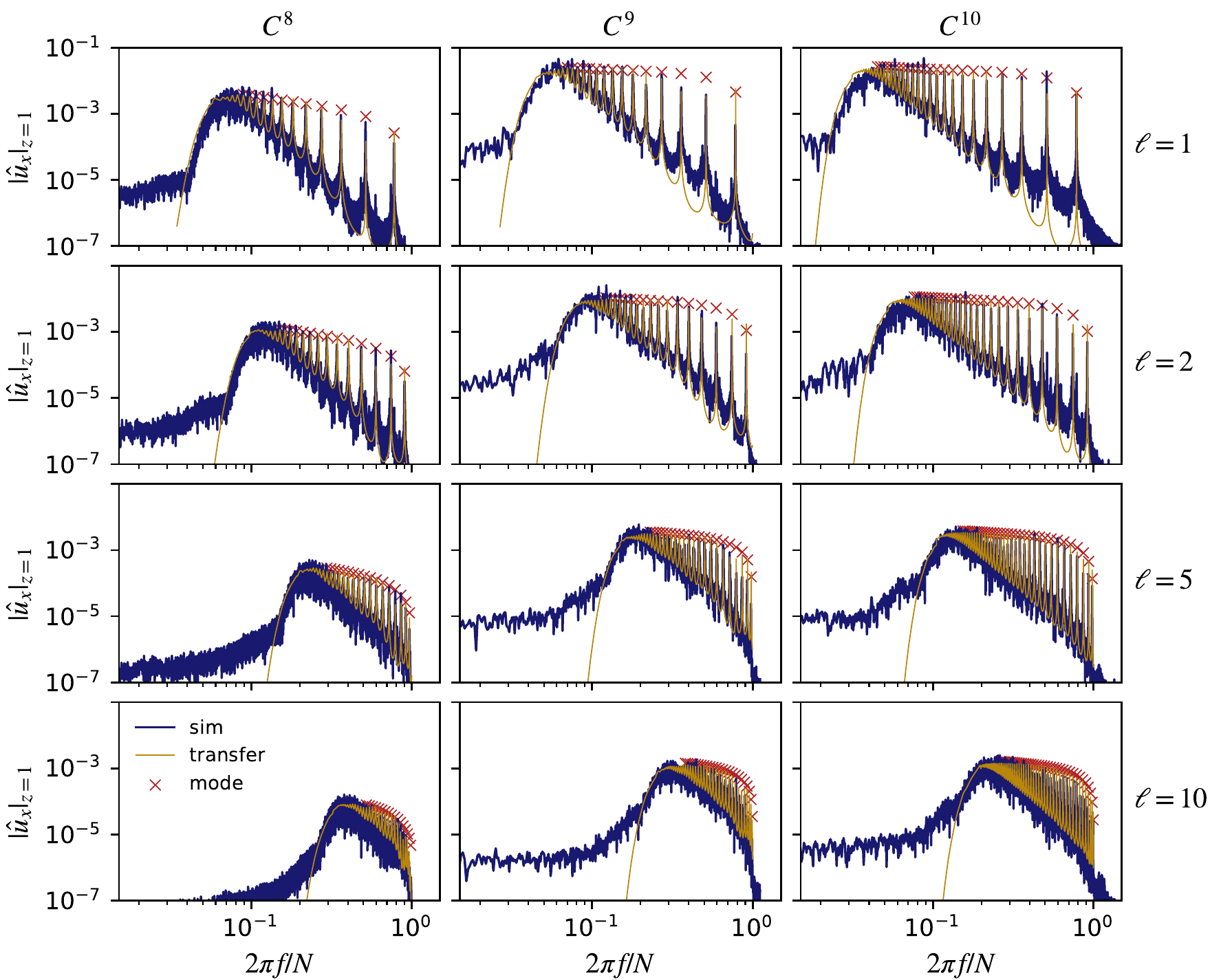}
  \caption{The frequency spectrum of the horizontal velocity at the surface for different horizontal wavenumbers, and in different simulations.
  We also plot the predicted spectrum derived using a transfer function (thin yellow, equation~\ref{eqn:transfer fit}), as well as predicted mode amplitudes (red crosses, equation~\ref{eqn:mode}).
  There is good agreement between the two theoretical predictions and the simulations.}
  \label{fig:spectrum_all}
\end{figure*}

\subsection{Mode Amplitudes}\label{sec:modes}

Instead of predicting the full wave spectrum, one can predict the amplitude of each of the peaks at the eigenfrequencies of the radiative zone.
\citet{Shiode2013} assumes that in a statistically steady state, the energy input into a mode by convection balances the energy dissipation of that mode.
They assume the energy input by convection is the wave energy flux, $(dF/d\ell df) \Delta \ell\, \Delta \omega/(2\pi)$.
Here $\Delta\ell$ and $\Delta \omega$ are the difference in $\ell$ and $\Re(\omega)$ between neighboring eigenmodes, and are related to the density of states.
The energy dissipation rate is $\gamma E_{\rm m}$, where $E_{\rm m}$ is the energy of the mode and $\gamma=\Im(\omega)$ is the dissipation rate of the mode.
It takes a time $\sim1/\gamma$ to reach this statistically steady state, which is longer than our simulation time for the highest-frequency and lowest-wavenumber modes.
For the surface amplitude, one uses the ratio
\begin{align}
R=\frac{|u_x|_{z=1}^2}{E_{\rm m}}=\frac{|u_x|_{z=1}^2}{\int |\vec{u}|^2 dz},
\end{align}
which can be calculated for each eigenmode derived in appendix~\ref{sec:eigenvalue}.

However, we find that this estimate does not match the simulation data.
That is because \citet{Shiode2013} did not take into account the finite width $\gamma$ of the peak associated with each mode.
To have a frequency-integrated energy of $E_{\rm tot}$, the maximum energy of each peak is $E_{\rm m}=E_{\rm tot} (\Delta\omega/\gamma)$.
This factor takes into account the density of states.
Then to balance energy injection from convection and energy dissipation, we have $\delta F = \gamma E_{\rm tot} = (\gamma^2/\Delta \omega) E_{\rm m}$.
We take $\Delta \omega$ to be the difference between the frequency of each mode and the mode with the next highest frequency (or $N$ for the highest-frequency mode), and only calculate the amplitudes of modes for which $\Delta \omega > 2\pi \gamma$.
Putting everything together, we estimate the mode amplitudes as
\begin{align}\label{eqn:mode}
|\hat{u}_x|_{z=1} = \mathcal{A}_M \left( \frac{\delta F  \Delta \omega}{\gamma^2} R\right)^{1/2},
\end{align}
where $\mathcal{A}_M$ is an overall amplitude we allow to fit each simulation.

\subsection{Comparison to Simulations}\label{sec:comparison}

We now compare the theoretical predictions, equations~\ref{eqn:transfer fit} \& \ref{eqn:mode}, to the results of our three simulations.
In each case, we use the wave flux spectrum defined in equation~\ref{eqn:fit} with parameters in table~\ref{tab:waveflux}.
We also picked amplitude factors $\mathcal{A}_T$ and $\mathcal{A}_M$ to improve the fit to the simulations (see table~\ref{tab:parameters}).
These uniformly scale the predictions up or down, and there is only one degree of freedom for all frequencies and wavenumbers.
Although the amplitude factors were chosen independently for each simulation, we find that $\mathcal{A}_M = 2.5 \mathcal{A}_T$.

\begin{table}
\centering
\begin{tabular}{ccc}\hline
Name & $\mathcal{A}_T$ & $\mathcal{A}_M$  \\ \hline \hline
$C^{8}$ & $0.4$ & $1$  \\
$C^{9}$ & $4$ & $10$ \\
$C^{10}$ & $1$ & $2.5$ \\
\end{tabular}
\caption{Amplitude factors $\mathcal{A}_T$ and $\mathcal{A}_M$ for the surface wave amplitude predictions, defined in equations~\ref{eqn:transfer fit} \& \ref{eqn:mode}.}\label{tab:parameters}
\end{table}

Figure~\ref{fig:spectrum_ell1} shows the surface frequency spectrum of the $\ell=1$ component of $u_x$, together with the predictions from the transfer function and the mode amplitude calculations.
The simulation, transfer function, and predicted mode amplitudes all have peaks at the same frequencies, and agree on the amplitudes of the peaks.
The agreement between the transfer function and mode amplitudes is particularly good.
The peaks in the simulation are sometimes higher than the theoretical predictions, sometimes lower.
This is not unexpected, as the waves are excited stochastically, so extremely long integrations are required to accurately determine the average mode amplitude.
The largest discrepancy is for the second-to-highest frequency mode with $2\pi f/N\approx 0.5$.
The transfer function is also able to reasonably reproduce the low-amplitude troughs in between the eigenfrequencies, although the agreement is worse for higher frequencies.
At low frequencies the wave amplitude decreases rapidly due to wave damping; this decay is well-captured by the transfer function, which is calculated via forced wave simulations in this regime.

In figure~\ref{fig:spectrum_all} we plot the surface frequency spectrum of $u_x$, together with the two theoretical models, for several different horizontal wavenumbers, and for all three simulations.
As in figure~\ref{fig:spectrum_ell1}, we find good agreement between the simulations and the predictions.
In simulations $C^8$ and $C^9$, the highest-frequency peaks are generally lower than predicted.
Also, at higher wavenumbers, the peaks of the spectrum are also often lower than predicted.
However, the amplitudes of the troughs between peaks seem to be well-predicted by the theory, with the possible exception of simulation $C^8$.
Overall, we find there is excellent agreement between the theoretical calculations and the simulations.
The theoretical predictions depend on only three parameters: the scaling law exponents $a$ and $b$, and the amplitude factor.
The scaling law exponents are already well-measured in three-dimensional Boussinesq simulations \citep{Couston2018}, and the amplitude factors (table~\ref{tab:parameters}) are all order unity.

\begin{figure}
  \centering
  \includegraphics[width=\linewidth]{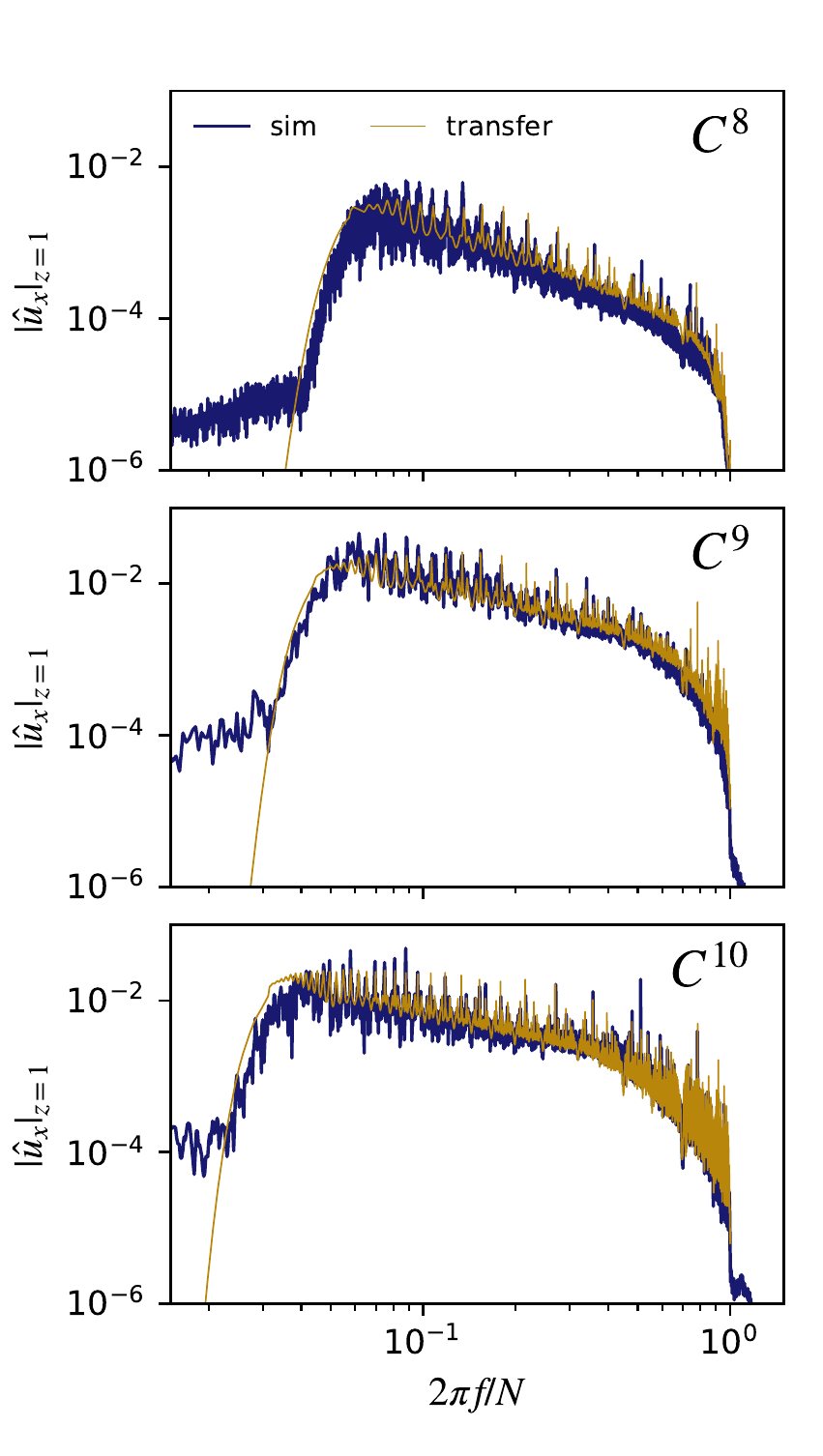}
  \caption{The surface frequency spectrum of the horizontal velocity.
  We filtered the horizontal velocity to only include horizontal wavenumbers $\ell \leq 20$, to more easily compare to the theoretical prediction.
  The blue line shows the simulation data, and the thin yellow line shows the theoretical prediction from the transfer function.
  The transfer function is able to reproduce most of the features of the full simulation.}
  \label{fig:spectrum_freq}
\end{figure}

Finally, in figure~\ref{fig:spectrum_freq} we plot the frequency spectrum of $u_x$ summed over horizontal modes up to $\ell=20$.
Because the transfer function is a good approximation to each individual $\ell$, it is no surprise that it agrees with the wavenumber-averaged spectrum.
Note however that the coefficients $a=3$ and $b=-13/2$ or $-15/2$ (table~\ref{tab:waveflux}) were chosen to improve the match between simulations and the transfer function calculation.
Both the simulations and the transfer function predictions show regularly-spaced peaks in their spectrum which are at the low-$\ell$ eigenfrequencies of the radiative zone.
They are not as obvious as when analyzing a single $\ell$ because the incoherent sum of all the other horizontal modes effectively raises the noise floor.
Here we only included the first 20 $\ell$ modes because they account for the horizontal velocity spectrum at low frequencies.
For $f\gtrsim 0.6N$, modes with $\ell>20$ raise the overall amplitude of the spectrum, but do not contribute any peaks.
One can compare the lower panel of figure~\ref{fig:spectrum_C10} to the lowest panel of figure~\ref{fig:spectrum_freq} to see the effect of modes with $\ell>20$.

\section{Summary}

In this work we presented simplified simulations to better understand the surface manifestation of internal gravity waves excited by convection.
We used the model of \citet{Couston2017} to run 2D Boussinesq simulations with a convection zone in the lower half of the domain, and a radiative zone in the upper half of the domain.
We first ran a series of simulations with a damping layer at the top of the radiative zone.
From these simulations, we measured the wave energy flux near the radiative-convective boundary.
The wave energy flux was in unexpectedly good agreement with theories of wave generation by 3D turbulence.
It is simple to measure the wave energy flux, as it does not require including the entire radiative zone, and the simulations can be run for a short time (e.g., tens of convection times, see \citealt{Couston2018}).

We then used two different theoretical approaches to translate the wave energy flux into a prediction of the surface manifestation of convective excited waves.
First, we calculated a transfer function, similar to \citet{Lecoanet2019}.
The transfer function relates the vertical velocity at the radiative-convective interface (which is given by the wave energy flux) to wave perturbations at the ``surface,'' or top, of the simulation.
Second, we calculated the amplitude of standing modes by assuming the energy input by convection matches the energy dissipation by diffusion \citep{Shiode2013}.
These both make theoretical predictions of the surface manifestation of the waves based off the wave flux.

To test these predictions, we ran a series of simulations with a reflecting top boundary, and measured the frequency spectrum of the internal gravity waves at the top of the simulation.
These simulations must be run for a long time (hundreds of convection times) before their surface spectra appear to saturate.
The wave generation is bursty, driven by intermittent intense convection events.
Despite this intermittency, we find excellent agreement between the surface frequency spectra in the simulations and the predicted spectra using the transfer function.
We found the original mode amplitude calculation had not correctly taken into account the finite frequency width of the peaks associated with each mode.
This effect increases the mode amplitudes relative to the predictions of \citet{Shiode2013}.
After taking into account the finite width of the peaks, we find good agreement with the simulations.

Our results show that using a transfer function is an accurate and efficient way to calculate the surface manifestation of convectively excited waves.
The transfer function can be calculated for a range of stellar models \citep[e.g.,][]{Lecoanet2019} if one makes an assumption about the spectrum of convectively excited waves.
Alternatively, one can run a short simulation including only part of the radiative zone to measure the wave flux, and then use the transfer function to determine the surface manifestation.

The transfer function calculation only works because the waves stay linear and because they can reflect off a top boundary.
Recently, \citet{Ratnasingam2020} ran two-dimensional simulations of internal gravity wave propagation in the radiative zone of intermediate-mass stars.
They found that nonlinear effects were weak, even when they excited waves using a spectrum with much greater energy than the spectrum we measure in our convection simulations.
We expect nonlinear effects to not be important \citep[see][for another argument based off observations]{Lecoanet2019}.

While our simulations had a reflecting top boundary, the near-surface layers of massive stars are convective \citep[e.g.,][]{Cantiello2019}.
It is unclear how or if internal gravity waves will be able to reflect off this upper convective boundary.
But note that the waves must reflect off the lower convective boundary in our simulations to generate the sharp peaks in the frequency spectrum.
While it is likely the surface convection will contribute some damping to the waves, it is unclear how important this is, as the waves are already strongly damped by radiative diffusion near the stellar surface.
Future simulations will be required to understand how the surface manifestation of waves are affected by surface convection.

Finally, we acknowledge the numerous physical effects that are important in stars but have been neglected in these simulations: rotation, magnetism, three-dimensionality, spherical geometry, density stratification, compressibility, etc.
Although these additional effects may add some technical complication in applying a transfer function to predict the surface manifestation of waves, we do not believe they will fundamentally limit the validity or utility of the approach.
Future work will explore to what extent these effects influence the generation, propagation, and surface manifestation of convectively excited internal gravity waves.

\section*{Acknowledgments}

\noindent{}We are grateful for useful discussions with Leo Horst, Philipp Edelmann, and Fritz R\"{o}pke that helped inspire this work.
We also thank Falk Herwig, Adam Jermyn, Anna Frishman, Geoff Vasil, Ben Brown, and Jeff Oishi for fruitful discussions.
DL is supported in part by NASA HTMS grant 80NSSC20K1280.
The Center for Computational Astrophysics at the Flatiron Institute is supported by the Simons Foundation.
Computations were conducted with support by the NASA High End Computing (HEC) Program through the NASA Advanced Supercomputing (NAS) Division at Ames Research Center on Pleiades with allocation GIDs s2276.
MB, BF and MLB acknowledge funding by the European Research Council under the European Union's Horizon 2020 research and innovation program through Grant No. 681835-FLUDYCO-ERC-2015-CoG.

\appendix

\section{Wave Flux in Simulations $D^8$ and $D^9$}\label{sec:waveflux89}

\begin{figure*}
  \centering
  \includegraphics[width=\linewidth]{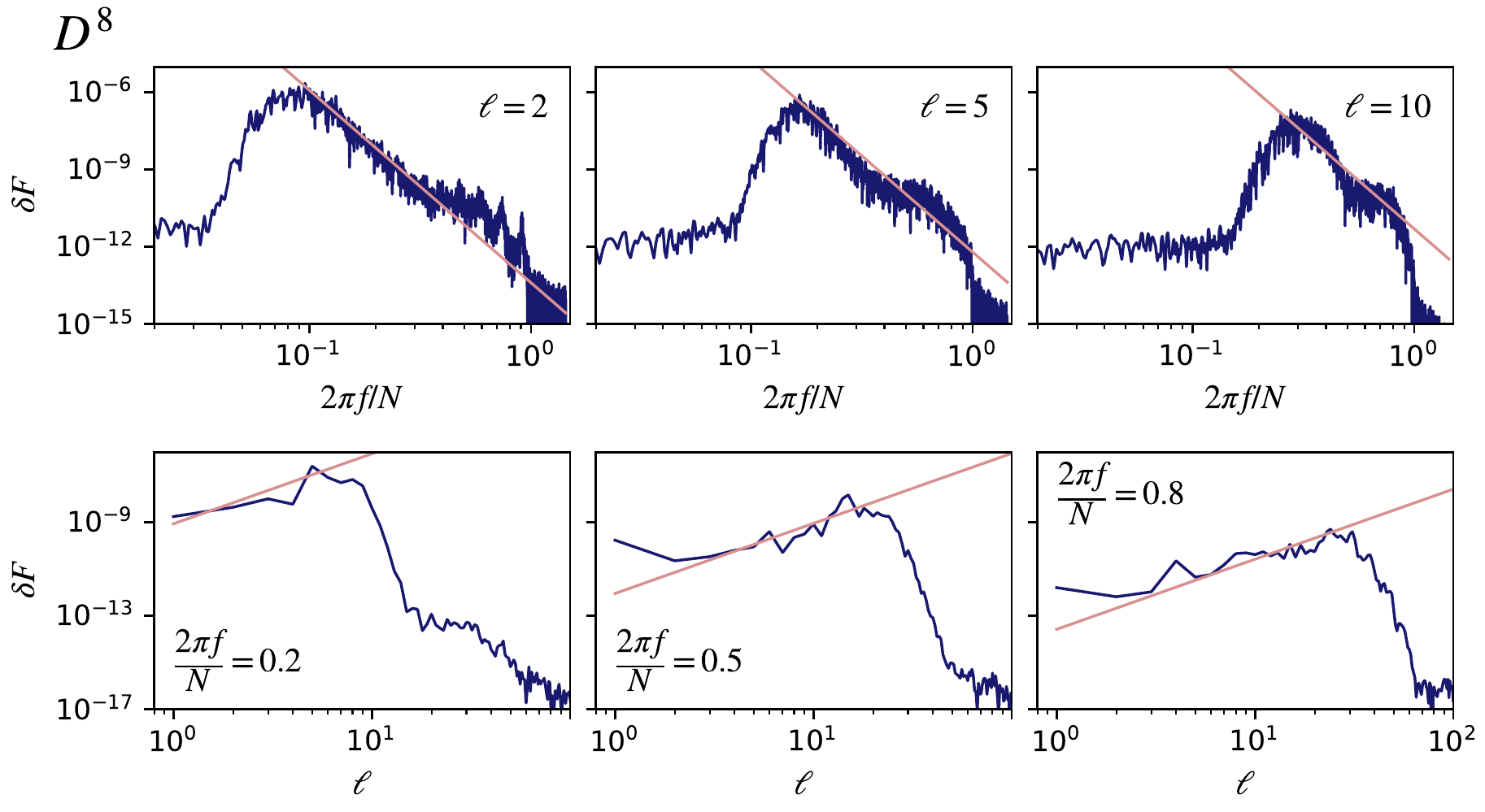}
  \caption{The wave flux spectrum as a function of $f$ (top row) and $\ell=k_x/(2\pi)$ (bottom row) for simulation $D^{8}$. The simulation data are plotted in the dark blue curve; the power-law expression ($f^{-15/2}\ell^{3}$) is plotted in the light red curve.}
  \label{fig:waveflux8}
\end{figure*}

\begin{figure*}
  \centering
  \includegraphics[width=\linewidth]{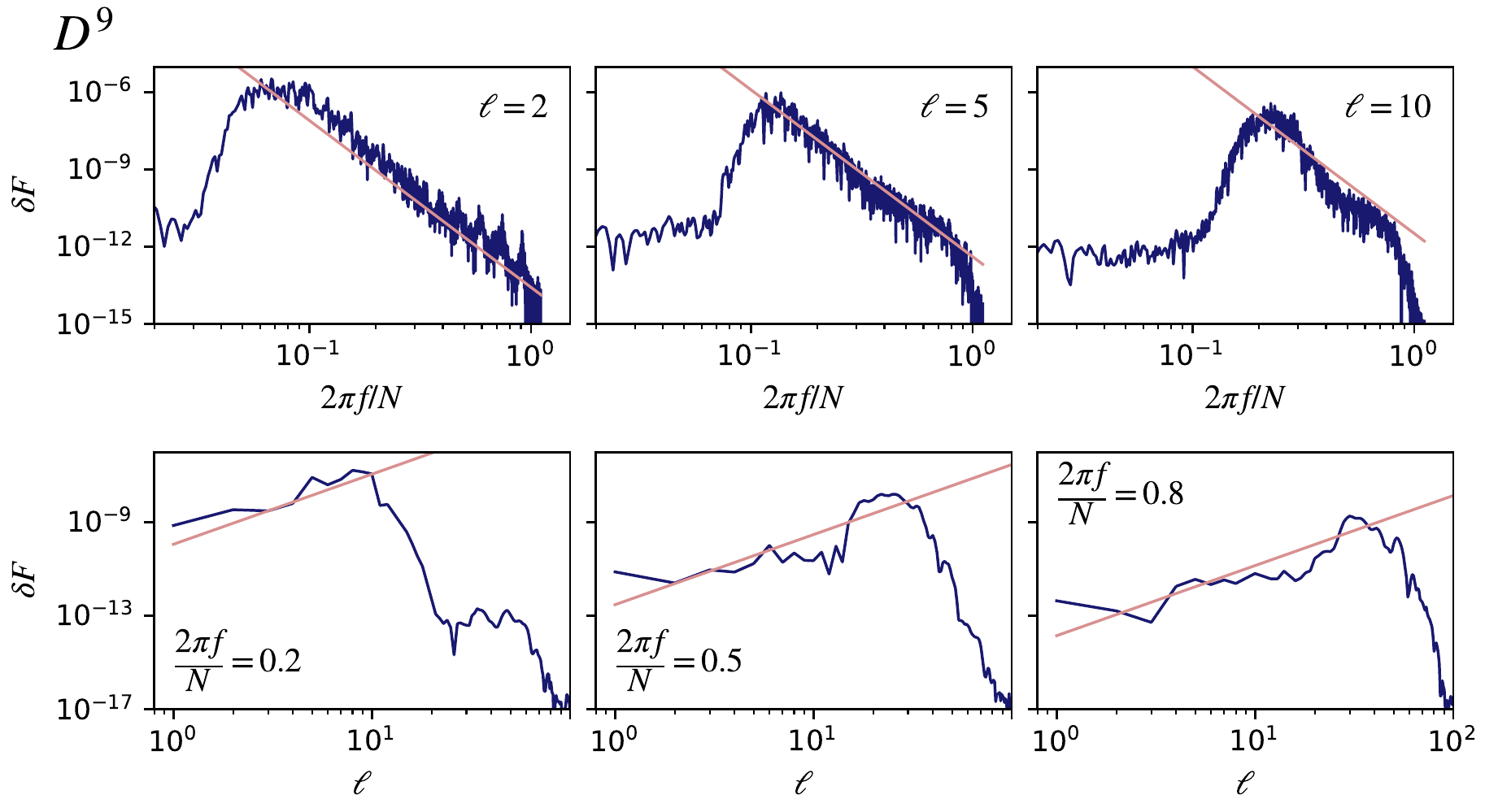}
  \caption{The wave flux spectrum as a function of $f$ (top row) and $\ell=k_x/(2\pi)$ (bottom row) for simulation $D^{9}$. The simulation data are plotted in the dark blue curve; the power-law expression ($f^{-13/2}\ell^{3}$) is plotted in the light red curve.}
  \label{fig:waveflux9}
\end{figure*}

Figures~\ref{fig:waveflux8} \& \ref{fig:waveflux9} show the wave flux spectra for simulations $D^8$ and $D^9$.
The power-law relation of equation~\ref{eqn:fit} seems to match best for simulation $D^9$.

\section{Eigenvalue Solves}\label{sec:eigenvalue}

To find the eigenmodes of the radiative zone, we compute the horizontal and temporal mean temperature as a function of $z$, $\overline{T}(z)$.
When linearizing the equations of motion (equations~\ref{eqn:momentum}--\ref{eqn:temperature}), we need the background temperature gradient.
We use $\partial_z \overline{T}$, but set this to zero below $z=0.45$ to avoid convection modes.
Assuming $T<0$ in the domain, we take $\delta \rho/\rho_0=\alpha S T$.
We use the Dedalus eigenvalue solver to compute the internal gravity wave eigenvalues and eigenfunctions.
We discretize $z$ using 128 Chebyshev modes between $z=0$ and $z=0.6$, and 256 Chebyshev modes between $z=0.6$ and $z=1$.
We first perform a dense eigenvalue solve.
To reject spurious modes, we then perform sparse eigenvalue solves in a higher-resolution domain with 192 modes in the lower part of the domain, and 384 modes in the upper part of the domain.
The sparse eigenvalue solve finds the eigenvalue closest to a target value; we use each of the eigenvalues of the dense solve as a target.
Modes are labeled as spurious if the fractional change in the eigenvalue is greater than $\epsilon=3\times 10^{-4}$, or if the pointwise difference in vertical velocity is greater than $\sqrt{\epsilon}$, when the vertical velocity is normalized to have a maximum amplitude of one.

\section{Direct Wave Forcing}\label{sec:forcing}

To calculate the wave transfer function at low frequencies we solved a forced, linearized wave equation.
That is, we solved the same equations as in the eigenvalue problem (including the modified $\partial_z \overline{T}$ profile to eliminate convection), but included an explicit forcing term.
The forcing term is
\begin{align}
\partial_t u_x + \ldots = \exp(i 2\pi f t) R(t) A(z; z_f),
\end{align}
with
\begin{align}
R(t) &=  \frac{1+\tanh\left[ (t-t_0)/\Delta t\right]}{2} \\
A(z; z_f) &= \frac{\tanh\left[ \frac{z-z_f+\delta z/2}{\delta z/10}\right] - \tanh\left[ \frac{z-z_f-\delta z/2}{\delta z/10}\right]}{2}. \\
\end{align}
The forcing term is centered at $z=z_f$, and we used $11$ values of $z_f$ between $0.5L$ and $0.55L$ with spacing $0.005L$.
We used $t_0=100/(2\pi f)$, $\Delta t = 10/(2\pi f)$, and $\delta z=0.01$.
The problem is discretized with $N_z$ Chebyshev modes between $z=0$ and $z=0.6$, and $N/2$ Chebyshev modes between $z=0.6$ and $z=1$.
For timestepping, we use a 2nd-order, two-stage, implicit-explicit Runge-Kutta scheme \citep{Ascher97}, where all linear terms are treated implicitly except the $u_z\partial_z \overline{T}$ term in the temperature equation, and the forcing term, which are treated explicitly.
We used a timestep size of $\approx 0.002\tau_c$.
We forced the system at 100 different frequencies, logarithmically spaced, between $f_{\rm min} \ell^{3/4}$ to $f_{\rm max} \ell^{3/4}$.
We use the $\ell^{3/4}$ scaling because the dissipation lengthscale of internal waves is $\ell_d^{-1}\sim \kappa \ell^3/f^4$.
The parameters used for each simulation are summarized in table~\ref{tab:forcing}.

\begin{table}
\centering
\begin{tabular}{ccccc}\hline
Name & $N_z$ & $2\pi f_{\rm min}/N$ & $2\pi f_{\rm max}/N$ & $2\pi f^*/N$\\ \hline \hline
$C^{8}$ & 256 & 0.057 & 0.11 & 0.11 \\
$C^{9}$ & 512 & 0.053 & 0.18 & 0.089 \\
$C^{10}$ & 512 & 0.038 & 0.094 &  0.063 \\
\end{tabular}
\caption{Numerical parameters for direct wave forcing simulations. $N_z$ is the number of Chebyshev modes used in the convection zone. $f_{\rm min}$ and $f_{\rm max}$ are related to the frequency range of the simulations.}\label{tab:forcing}
\end{table}

The simulations are run for $10^{5}$ timesteps.
We measure the average amplitude of $u_x$ at $z=1$ for the final 15\% of the simulation.
This is more than enough time for the simulations to reach a statistically steady state in this low frequency range.
When calculating the transfer function, we must decide for which frequencies to use the eigenfunction expansion, and for which frequencies to use the direct forcing simulations.
The two approaches give similar results at intermediate frequencies, and we transition between the two at $f^*$, which is reported for each simulation in table~\ref{tab:forcing}.

\bibliographystyle{mn2e}
\bibliography{waves}

\end{document}